\RequirePackage{fix-cm}
\documentclass[smallextended, draft, fleqn]{svjour3}       
\smartqed  
\usepackage{graphicx}
 \usepackage{mathptmx}      
\usepackage[utf8]{inputenc}
\usepackage{geometry}

\usepackage[intlimits, sumlimits]{amsmath}
\usepackage{amssymb}
\usepackage{amsbsy}  
\usepackage{graphicx}

\usepackage{nicefrac}

\renewcommand{\bar}[1]{\overline{#1}}
\renewcommand{\hat}[1]{\widehat{#1}}

\def\DD{\mathcal{D}}

\newcommand{\ord}[1]{\mathcal{O}\left( #1 \right)}

\def\dd{\text{d}}
\newcommand{\bb}[1]{\mathbf{#1}}
\def\pd{\partial}

\def\eps{\varepsilon}

\def\lagr{{\cal L}}
\def\fii{\varphi}
\def\scri{{\cal I}}
\def\FF{\mathcal{F}}

\newcommand{\Yslm}[3]{{}_{#1}Y_{{#2#3}}}
\newcommand{\Ylm}[2]{Y_{{#1#2}}}

\def\DD{\mathcal{D}}

\def\shear{\sigma^{(0)}}
\def\shearcc{\bar{\sigma}{}^{(0)}}
\def\sheardot{\dot{\sigma}{}^{(0)}}
\def\shearccdot{\dot{\bar{\sigma}}{}^{(0)}}
\def\phizero{\phi^{(0)}}
\def\phizerocc{\bar{\phi}{}^{(0)}}

\begin{document}

\title{On the Bondi mass of Maxwell-Klein-Gordon spacetimes}



\author{Martin Scholtz         \and
        Luk\'a\v{s} Holka 
}


\institute{M. Scholtz\at Department of Applied Mathematics, Na Florenci 25, Prague, Czech Republic, \\
\email{scholtz@fd.cvut.cz}
\and
L. Holka \at
Institute of Theoretical Physics, V Hole\v{s}ovi\v{c}k\'ach 2, Prague, Czech Republic, \\
\email{lukas.holka@gmail.com}
}

\date{Received: 27 September 2013 / Accepted: 5 January 2014}

\maketitle

\begin{abstract}
In this paper we calculate the Bondi mass of asymptotically flat spacetimes with interacting electromagnetic and scalar fields. The system of coupled Einstein-Maxwel-Klein-Gordon equations is investigated and corresponding field equations are written in the spinor form and in the Newman-Penrose formalism. Asymptotically flat solution of the resulting system is found near null infinity. Finally we use the asymptotic twistor equation to find the Bondi mass of the spacetime and derive the Bondi mass-loss formula. We compare the results with our previous work \cite{BST2} and show that, unlike the conformal scalar field, the (Maxwell-)Klein-Gordon field has negatively semi-definite mass-loss formula.


\end{abstract}

\section{Introduction}
\label{sec:intro}

It is a well known fact that the energy-momentum of gravitational field cannot be introduced at the {\itshape local} level which is, after all, the consequence of the equivalence principle. Since it is highly desirable to have a meaningful notion of the energy and the momentum, many suggestions have been made in order to define the {\itshape quasi-local} energy-momentum which is associated with, e.g., a compact spacelike hypersurface $\Sigma$ with boundary $S$, rather than with a spacetime point. The quasi-local quantities are usually expressed as the surface integrals over the 2-surface $S$. The most influential suggestions are, for example, those of Penrose \cite{PenroseMass}, Hawking \cite{HawkingMass}, Dougan and Mason \cite{DouganMasonMass} and Brown and York\cite{BrownYorkMass}. For extensive reviews on the subject, see \cite{SzabadosReview,JaramilloGourghoullon}. 

On the other hand, in the case of asymptotically flat spacetimes there is a well-defined notion of {\itshape global} energy-momentum associated with the entire spacetime (ADM mass \cite{ADMmass} defined at spatial infinity) or energy-momentum associated with an isolated gravitating source (Bondi mass \cite{BondiBurgMetzner} defined at null infinity). Hence, one of the natural criteria of the plausibility of particular quasi-local energy-momentum is whether it coincides with the ADM mass or the Bondi mass in the limit of the large spheres near spatial or null infinity \cite{SzabadosReview}. 

Standard expression for the Bondi energy-momentum of electro-vacuum spacetimes in the Newman-Penrose formalism has the form
\[
P^{\bb{AA'}} = - \oint_S \left( \Psi_2^{(0)} + \shear\,\shearccdot\right)\,\omega^{\bb{A}}_0\,\bar{\omega}^{\bb{A'}}_0\,\dd S,
\]
where $\Psi_2^{(0)}$ is the leading $\ord{r^{-3}}$ term in the asymptotic expansion of the $\Psi_2$-component of the Weyl spinor, $\shear$ is the asymptotic shear of Newman and Penrose, $\omega^{\bb{0}}_A$ and $\omega^{\bb{1}}_A$ are asymptotic spinors \cite{SzabadosReview} and $\omega^{\bb{A}}_0 = \omega^{\bb{A}}_A o^A$, where $o^A$ is the element of GHP spinor dyad \cite{GHP}. The dot means the derivative with respect to (retarded) time $u$. In \cite{BST2} we have shown that this result remains true for the spacetimes with conformally invariant scalar field sources. In the presence of the massless Klein-Gordon scalar field, however, the scalar field contributes to the Bondi energy and the correct expression for the Bondi mass (energy) is (in the conventions used in this paper)
\begin{align}
M_B &= - \frac{1}{2\sqrt{\pi}}\oint_S \left( \Psi_2^{(0)} + \shear\,\shearccdot + \frac{1}{6}\pd_u (\phizero\,\phizerocc)\right)\,\dd S,
\label{eq:Bondi mass BST2}
\end{align}
where $\phizero$ is now the leading $\ord{r^{-1}}$ term in the asymptotic expansion of the scalar field.

A crucial property of the Bondi energy is that it should decrease whenever the system emits gravitational (or another) radiation. As we have shown in \cite{BST2}, in the case of massless Klein-Gordon field the mass-loss formula acquires the form
\begin{align}
 \dot{M}_B = - \frac{1}{2\sqrt{\pi}} \oint \left(
 \sheardot\,\shearccdot + \dot{\phi}{}^{(0)}\,\dot{\bar{\phi}}{}^{(0)}\right)\,\dd S,
 \label{eq:Bondi massloss BST2}
\end{align}
so that the Bondi mass is a non-increasing function of time $u$. For the conformally invariant scalar field, resulting ``mass-loss'' formula is indefinite and reads
\begin{align}
\dot{M}_B = - \frac{1}{2\sqrt{\pi}}\oint_S \left( \sheardot\,\shearccdot +
2 (\dot{\phi}{}^{(0)})^2 - \phizero\,\ddot{\phi}{}^{(0)}\right)\,\dd S.
\label{eq:Bondi massloss indefinite}
\end{align}
Hence, in this case the Bondi mass is not a monotonic function of time, which can be traced back to the fact that the energy-momentum tensor for the conformally invariant scalar field does not obey the energy condition $T_{ab} l^a n^b \geq 0$ for any future null vectors $l^a$ and $n^a$.

In this paper we investigate the natural generalization of these calculations and we calculate the Bondi mass of the spacetimes with interacting electromagnetic and scalar fields. The purpose is twofold. It seems that the analysis of the Bondi mass of Maxwell-Klein-Gordon spacetimes in the Newman-Penrose formalism is missing (see, however, \cite{ChruscielJezierskiKijowski,HuangCY} for some results on the scalar field in the Hamiltonian formalism). Hence, our first goal is to fill this gap.

The Penrose mass has been calculated for a wide class of spacetimes in \cite{TodPenroseMass,TodExamples,TodMore}, but the spacetimes with scalar field sources are not included. In fact, only a very few exact solutions of coupled Einstein-Maxwell-Klein-Gordon equations are known, e.g. \cite{Gegenberg}. On the other hand, there is a chance that at least some properties of the Penrose mass can be understood without having an exact solution. The idea is to apply standard 3+1 decomposition of the spacetime with electromagnetic and scalar field sources and analyse the constraints which must be satisfied on the initial Cauchy hypersurface. The 2-surface $S$ can be chosen to lie in this initial hypersurface and one can hope that the constraints will be easier to solve than the full set of equations. In this context, the present paper is a preliminary work: the Penrose mass calculated by the analysis sketched in this paragraph can be examined to have the correct large sphere limit.

The paper is organized as follows. In the section \ref{sec:field eqs} we introduce standard equations governing the system of coupled gravitational, electromagnetic and scalar fields and translate them into the spinor formalism. In the appendix \ref{app:projections} we present the Newman-Penrose projections of these equations. Next we consider an asymptotically flat spacetime with the electromagnetic and scalar field sources which is analytic at the future null infinity $\scri^+$. The asymptotic behaviour of the Newman-Penrose quantities describing the gravitational, scalar and electromagnetic fields is investigated in the section \ref{sec:asymptotics}. In the next section \ref{sec:solution} we present the asymptotic solution of Einstein-Maxwell-Klein-Gordon equations and finally in the section \ref{sec:bondi} we calculate the Bondi mass of the spacetime and find corresponding mass-loss formula which is presented both in terms of the four-potential and in the gauge invariant form.


\section{Field equations}
\label{sec:field eqs}

In this section we introduce field equations of interacting electromagnetic, scalar and gravitational fields in the spinor form. Resulting system of equations will be referred to as the {\itshape Einstein-Maxwell-Klein-Gordon equations} and corresponding spacetime will be called {\itshape electro-scalar spacetime} for the sake of brevity.

The gauge invariant Lagrangian of the coupled scalar and electromagnetic fields can be written in the form \cite{HawkingEllis}
\begin{align}
\lagr = (\DD_a\phi)(\DD^a\bar{\phi}) - m^2\,\phi\,\bar{\phi} - \frac{1}{4}\,F^{ab}\,F_{ab} \, ,
\label{eq:lag SEM}
\end{align}
where $\phi$ is a scalar field with charge $e$, $\bar{\phi}$ its complex conjugate with charge $-e$, $m$ is the mass of the scalar field and $F_{ab}$ is standard Faraday 2-form. When acting on the uncharged fields, the gauge covariant derivative $\DD_a$ coincides with the usual covariant derivative $\nabla_a$, otherwise its action on an arbitrary tensor field $T^{a..b}_{c..d}$ with the charge $e$ is defined by \cite{PenroseRindlerI}
\begin{align}
\DD_f T^{a..b}_{c..d} = \nabla_f T^{a..b}_{c..d} + i \, e\, A_f\,T^{a..b}_{c..d},
\label{eq:gauge cov der}
\end{align} 
with $A_a$ being the four-potential. The Lagrangian (\ref{eq:lag SEM}) yields, through the standard Euler-Lagrange equations, familiar field equations
\begin{align}
\left( \DD_a \DD^a + m^2\right) \phi = 0, \qquad
\left( \DD_a \DD^a + m^2\right) \bar{\phi} &= 0, \qquad
\nabla^a F_{ab} = i\, e\left( \bar{\phi} \DD_b \phi  - \phi \DD_b\bar{\phi}\right).
\label{eq:eqs of motion}
\end{align}

Next we wish to rewrite these equations as a system of first-order spinorial equations. Electromagnetic spinor $\phi_{AB}$ is related to the potential $A_a$ by
\begin{align}
 \phi_{AB} = \nabla_{X'(A}A^{X'}_{B)}.
 \label{eq:phi_AB curl}
\end{align}
We reduce the gauge freedom imposing standard Lorenz condition $\nabla_a A^a=0$, so that the equation (\ref{eq:phi_AB curl}) simplifies to
\begin{align}
\nabla^{A'}_A A_{BA'} = -\,\phi_{AB}\,.
\label{eq:nabla A}
\end{align}
Because we prefer our equations to be of the first order, we retain both $\phi_{AB}$ and $A_{AA'}$ in future formulae and equation \eqref{eq:nabla A} will be regarded as a dynamical equation for the potential $A_a$. Spinor form of \eqref{eq:eqs of motion} then implies the equation for $\phi_{AB}$:
\begin{align}
 \nabla^A_{B'} \phi_{AB} = \frac{ie}{2}\left( \bar{\phi}\DD_b\phi - \phi \DD_b \bar{\phi}\right)
=
\frac{ie}{2}\left(
\bar{\phi}\,\fii_b - \phi\,\bar{\fii}_b
\right) - e^2\,\phi\,\bar{\phi}\,A_b.
\label{eq:nabla phiAB}
\end{align}

In order to derive first-order equations for the scalar field, we introduce notation (cf. \cite{BST2})
\begin{align}
 \fii_a = \nabla_a \phi \quad \text{and} \quad \fii_{AA'} = \nabla_{AA'} \phi
\end{align}
which eliminates formally the second derivatives of the scalar field $\phi$ that are present in the equations \eqref{eq:eqs of motion}. These equations are equivalent to the wave equation
\begin{align}
 \Box \phi &=- 2\,i\,e\,A^a\,\fii_a + \left(e^2\,A^a\,A_a - m^2\right)\phi 
\label{eq:dAlembert phi}
\end{align}
and its complex conjugate. At this point we could employ the Newman-Penrose formalism and express $\Box \phi$ with the help of only the first derivatives of $\fii_a$ and the spin coefficients. However, it is more convenient to decompose spinor $\nabla_{A'}^A\fii_{AB'}$ into its symmetric and antisymmetric parts,
\[
 \nabla^A_{A'} \fii_{AB'} = \nabla^A_{(A'}\fii_{B')A} + \frac{1}{2}\,\epsilon_{A'B'}\,\nabla_{X'}^A\fii_A^{X'} = 
- \Box_{A'B'}\phi - \frac{1}{2}\,\epsilon_{A'B'}\,\Box\phi,
\]
and use $\Box_{A'B'}\phi\, =\, 0$. (Commutator $\Box_{AB} =  \nabla_{X'(A}\nabla^{X'}_{B)}$ annihilates scalar quantities.) Now the scalar equation \eqref{eq:dAlembert phi} is equivalent to the spinor equation
\begin{align}
\nabla^A_{A'}\fii_{AB'} &= \,i\,e\,A^c\fii_c\,\epsilon_{A'B'} + \frac{1}{2}\left(m^2 - e^2\,A^c A_c\right)\phi\,\epsilon_{A'B'}.
\label{eq:nabla fii}
\end{align}
If, on the other hand, we apply the procedure of spinor decomposition to covariant derivatives $\DD_a =\DD_{AA'}$ in the Klein-Gordon equation \eqref{eq:eqs of motion}, we arrive at somewhat more elegant formula
\begin{align}
\DD_A^{X'}\DD_{B X'} \phi &= \frac{1}{2}\,m^2\,\phi\,\epsilon_{AB}  - i\,e\,\phi\,\phi_{AB}.
\label{eq:KG spinor}
\end{align}
Here, the Lorenz condition {\itshape has not} been imposed and equation \eqref{eq:KG spinor} is manifestly gauge-invariant.

Now we turn our attention to equations of gravitational field which is described by the Newman-Penrose spin coefficients, the Weyl spinor $\Psi_{ABCD}$, the Ricci spinor $\Phi_{ABA'B'}$ and the scalar curvature $\Lambda=  R/24$. Equations for the spin coefficients follow from the spinorial form of the Ricci identities \cite{StewartBook}
\begin{align}
\Box_{CD} \xi_A = \Psi_{ABCD}\,\xi^B - 2 \Lambda\,\epsilon_{A(C}\,\xi_{D)}, \qquad
\Box_{C'D'} \xi_{A} = \Phi_{ABC'D'} \,\xi^B,
\end{align}
where $\xi_A$ is chosen to be one of the basis spinors $o_A$ and $\iota_A$. The Weyl spinor and the Ricci spinor satisfy the Bianchi identities
\begin{align}
\nabla_{A'}^D \Psi_{ABCD} = \nabla^{B'}_{(A} \Phi_{BC)A'B'}, \qquad
\nabla^{BB'} \Phi_{ABA'B'} = - 3\,\nabla_{AA'} \Lambda.
\label{eq:Bianchi ids spinor}
\end{align}
Moreover, the Ricci spinor and the scalar curvature are related to the energy-momentum tensor by the Einstein equations \cite{BST1}
\begin{align}
\Phi_{ABA'B'} =  4\pi \, T_{(AB)(A'B')}, \qquad
3\Lambda  =  \pi \, T_{XY'}{}^{XY'}.
\label{eq:Einsteins eqs}
\end{align}  
In order to obtain the energy-momentum tensor $T_{ab}$ we vary the action of the electro-scalar field with the Lagrangian (\ref{eq:lag SEM}) with respect to the metric $g^{ab}$. This yields (cf. \cite{HawkingEllis})
\begin{multline}
T_{ab}  =\frac{1}{4\pi} \left[
\left(\DD_{(a}\phi\right)\left(\DD_{b)}\bar{\phi}\right) - \frac{1}{2}\,F_{ac}\,F_b^{\phantom{b}c} - \frac{1}{2}\,g_{ab}\,\lagr
\right] \\
=\frac{1}{4\pi} \left[
\left(\DD_{(a}\phi\right)\left(\DD_{b)}\bar{\phi}\right) + \phi_{AB}\,\bar{\phi}_{A'B'} - 
\frac{1}{2}\,g_{ab}\left(\DD_c\phi\right)\left(\DD^c\bar{\phi}\right) + \frac{1}{2}\,m^2\,g_{ab}\,\phi\bar{\phi}\right].
\label{eq:energy-momentum tensor}
\end{multline}
where the factor $(4\pi)^{-1}$ has been included for convenience. Using the Einstein equations \eqref{eq:Einsteins eqs} we find that the Ricci spinor and the scalar curvature are given by relations
\begin{align}
 \Phi_{ABA'B'}=  \left(\DD_{(A(A'}\phi\right)\left(\DD_{B')B)}\bar{\phi}\right) + \phi_{AB}\,\bar{\phi}_{A'B'}, 
 \qquad
 \Lambda = \frac{1}{12}\left[ - \left(\DD_a\phi\right)\left(\DD^a\bar{\phi}\right) + 2\,m^2\,\phi\,\bar{\phi}\right].
\label{eq:Phi Lambda}
\end{align}

To summarize, the unknown variables representing the matter fields are the potential $A_a$ governed by equation \eqref{eq:nabla A}, the electromagnetic spinor $\phi_{AB}$ governed by \eqref{eq:nabla phiAB} and the scalar field satisfying \eqref{eq:KG spinor}. Corresponding Newman-Penrose projections are summarized in the appendix \ref{app:projections}, equations \eqref{eq:Nabla A NP}, \eqref{eq:Maxwell eqs} and \eqref{eq:Nabla fii}. The components of the Ricci spinor and the scalar curvature are given by \eqref{eq:Phi Lambda} and their are listed explicitly in the Newman-Penrose form in the appendix \ref{app:projections}, equations \eqref{eq:Ricci comps} and \eqref{eq:Lambda NP}. The Weyl spinor and the Ricci spinor satisfy the Bianchi identities \eqref{eq:Bianchi ids spinor}. Corresponding Newman-Penrose equations \cite{PenroseRindlerI,StewartBook} are listed in the appendix for the reference purposes.


\section{Asymptotic behaviour of the fields}
\label{sec:asymptotics}

We are interested in a weakly asymptotically simple solution of the Einstein-Maxwell-Klein-Gordon equations which is analytic\footnote{In order to calculate the Bondi mass, the analyticity is not necessary and weaker assumptions on the differentiability of the solution could be imposed. In what follows we use the analyticity to argue that the mass of the Klein-Gordon field must be zero.} in the neighbourhood of the future null infinity $\scri^+$. We employ the notation $(\hat{M}, \hat{g}_{ab})$  for the unphysical spacetime and $(M, g_{ab})$ for the physical one, where, by assumption of weak asymptotic simplicity, the two metrics are related by conformal rescaling  
\begin{align}
\hat{g}_{ab} = \Omega^2 g_{ab}.
\label{eq:ConRes}
\end{align}

To proceed further we need to establish a coordinate system and the Newman-Penrose null tetrad in a neighbourhood of $\scri^+$. In accordance with \cite{StewartBook} we introduce coordinates $x^\mu = (u, r, x^2, x^3)$, where $x^I,  I = 2,\,3$, are arbitrary coordinates on the 2-sphere, $u$ is an affine parameter along null generators of $\scri^+$ and $r$ is an affine parameter along null hypersurfaces intersecting $\scri^+$ in cuts $u = \text{constant}$. Vector $l^a$ is chosen to be tangent to these null hypersurfaces and orthogonal to the cuts of constant (both) $u$ and $r$. Null vectors $m^a$ and $\bar{m}^a$ are chosen so as to span the tangent space of these cuts. Resulting null tetrad has the following properties.

\begin{itemize}
	\item $l^a$ and $n^a$ are real and null vectors normalized by  $l^a n_a\, = \, 1$. Vector $m^a$ and its complex conjugate $\bar{m}^a$ are null and complex, satisfying the condition $m^a \bar{m}_a\, = \, -1$. Remaining scalar products between these four vectors are all zero. Their components with respect to the basis induced by the coordinates $(u, r, x^2, x^3)$ read
	\begin{align}       
                l^\mu  =  (0, 1, 0, 0),  \quad n^\mu  =  (1, H, C^2, C^3), \quad
                m^\mu  =  (0, 0, P^2, P^3), \quad\bar{m}^\mu  =  (0, 0, \bar{P}^2, \bar{P}^3).
       \label{eq:NP tetrad coords}
  \end{align}
  \item There exists a spin basis $(o^A,\iota^A)$ such that
   \begin{align}       
                l^a  =  o^A\bar{o}^{A'}, \quad  n^a  =  \iota^A\bar{\iota}^{A'}, \quad
                m^a  =  o^A \bar{\iota}^{A'}, \quad \bar{m}^a  =  \iota^A\bar{o}^{A'}.
   \end{align}
  \item Functions $H$, $C^I$ and $P^I$ are subject to the frame equations :
      \begin{subequations}
                 \begin{align}
                   D H &= - \,\gamma - \bar{\gamma}, \label{eq:DH} \\
                   D C^I &=  \,2\pi P^I + 2\bar{\pi}\bar{P}^I, \label{eq:DCI}\\
                   D P^I &= \rho P^I + \sigma \bar{P}^I, \label{eq:DPI}\\
                   \Delta P^I  - \delta C^I &= (\gamma-\bar{\gamma}-\mu) P^I - \bar{\lambda} \bar{P}^I, \label{eq:DeltaPI - deltaCI}\\
                   \delta H &= - \bar{\nu}, \label{eq:deltaH}\\
                   \bar{\delta} P^I - \delta\bar{P}^I &= (\alpha-\bar{\beta}) P^I + (\beta-\bar{\alpha})\bar{P}^I \label{eq:bardeltaPI-deltabarPI},
                  \end{align}
                  \label{eq:frame equations}
      \end{subequations}
      \noindent where we have used the standard Newman-Penrose notation
      \begin{align}
      l^a\nabla_a = D, \quad  n^a\nabla_a = \Delta, \quad 
      m^a\nabla_a = \delta, \quad \bar{m}^a\nabla_a = \bar{\delta}.
      \end{align}

    \item Some of the spin coefficients get simplified:
         \begin{align}
                 \epsilon =  0, \quad \kappa =  0, \quad
                 \mu = \bar{\mu}, \quad \rho = \bar{\rho}, \quad
                 \bar{\pi} = \tau = \bar{\alpha}+\beta. \label{eq:alpha beta pi tau}
         \end{align}

     \item In accordance with (\ref{eq:ConRes}) we choose a spin basis in the unphysical spacetime
          \begin{align}
                         \hat{o}^A = \Omega^{-1}\,o^A, \quad
                         \hat{\iota}^A = \iota^A, \quad
                         \hat{o}_A = o_A, \quad
                         \hat{\iota}_A = \Omega\,\iota_A.
                         \label{eq:spin basis rescaling}
         \end{align} 
         Associated unphysical null tetrad then reads
                \begin{align}
                  \hat{l}^a = \Omega^{-2}\,l^a, \quad
                  \hat{n}^a = n^a, \quad
                  \hat{m}^a = \Omega^{-1}\,m^a.
                \label{eq:NP tetrad rescaling}
                \end{align}
          \noindent We assume that unphysical spinors $\hat{o}^A$ and $\hat{\iota}^A$ are regular on $\scri^+$ which implies that physical spinor $o^A = \Omega \hat{o}^A$ vanishes on $\scri^+$ while the spinor $\iota^A$ remains non-vanishing there. 
          
     \item In the neighbourhood of $\scri^+$ we can use the conformal factor $\Omega$ as a coordinate instead of $r$ by setting $\dd\Omega /\dd r= - \Omega^2$. The Newman-Penrose operators (acting on scalars) then read
             \begin{align}
                 D = -\Omega^2\pd_\Omega,\quad
                 \Delta = \pd_u - \Omega^2\,H\,\pd_\Omega + C^I\pd_I, \quad
                 \delta = P^I\pd_I.
             \label{eq:NP operators asymp}
             \end{align}
    \noindent In particular, we have 
            \begin{align}
                    D\Omega = -\,\Omega^2,\quad
                    \Delta \Omega = -\,\Omega^2\,H,\quad
                    \delta\Omega = \bar{\delta}\Omega = 0.
            \label{eq:ders of Omega}
            \end{align}
    \noindent In addition, by (\ref{eq:spin basis rescaling}) we have
              \begin{align}
                C^I = \ord{\Omega}, \quad
                P^I = \ord{\Omega}.
              \label{eq:CI PI asymp}
              \end{align}
            
\end{itemize}

Next we establish the asymptotic behaviour of the spin coefficients under the assumption that unphysical spin coefficients are regular on $\scri^+$, i.e. they are of order $\ord{1}$. Under the conformal rescaling, the spin coefficients transform as
\begin{align}
\kappa &= \Omega^3\,\hat{\kappa}, &   \tau &= \Omega\,\hat{\tau} + \hat{\delta}\Omega,   &
\sigma &= \Omega^2\,\hat{\sigma}, &   \rho &= \Omega^2\,\hat{\rho} + \Omega\,\hat{D}\Omega, 
\nonumber \\
\eps &= \Omega^2\,\hat{\eps},   &  \gamma &= \hat{\gamma} + \Omega^{-1}\hat{\Delta}\Omega,  &
\beta &= \Omega\,\hat{\beta},   &  \alpha &= \Omega\,\hat{\alpha} + \hat{\bar{\delta}}\Omega,
\label{eq:spin coeffs rescaling}\\
\pi &= \Omega\,\hat{\pi} - \hat{\bar{\delta}}\Omega,  & \nu &= \Omega^{-1}\,\hat{\nu},  &
\mu &= \hat{\mu} - \Omega^{-1}\,\hat{\Delta}\Omega,   & \lambda &= \hat{\lambda}. \nonumber
\end{align}
These relations have been derived using the definitions of spin coefficients, the rule for the transformation of the covariant derivative \cite{PenroseRindlerII,StewartBook} and the behaviour of the spin basis (\ref{eq:spin basis rescaling}). Derivatives with the hats are operators associated with the unphysical spin basis $\hat{o}^A$ and $\hat{\iota}^A$. We assume the order $\ord{1}$ for all unphysical quantities. 

In the tetrad introduced above, coefficients $\eps$ and $\kappa$ vanish and thus, by (\ref{eq:spin coeffs rescaling}), their unphysical counterparts $\hat{\eps}$ and $\hat{\kappa}$ vanish as well. Moreover, by (\ref{eq:ders of Omega}) we have
\[
 \tau = \Omega\hat{\tau} = \ord{\Omega}, \quad
 \pi = \Omega\hat{\pi} = \ord{\Omega}, \quad
 \alpha = \Omega\hat{\alpha} = \ord{\Omega}, \quad
 \beta = \Omega\hat{\beta}  = \ord{\Omega}.
\]
For the coefficients $\gamma, \mu$ and $\lambda$ we find
\begin{align}
 \gamma = \hat{\gamma} - \Omega\,H = \ord{1}, \quad
 \mu = \hat{\mu} + \Omega\,H = \ord{1}, \quad
 \lambda=\hat{\lambda} = \ord{1}.
\end{align} 
The coefficient $\nu$ is apparently divergent on $\scri^-$, 
\begin{align}
 \nu &= \Omega^{-1}\,\hat{\nu} = \ord{\Omega^{-1}},
\end{align}
because of \eqref{eq:spin basis rescaling}, but we will show that in fact $\nu=\ord{\Omega^2}$. The coefficient $\sigma$ is of the order
\begin{align}
 \sigma &= \Omega^2\,\hat{\sigma} = \ord{\Omega^2}.
\end{align}
Finally, for the coefficient $\rho$ we have (see \cite{StewartBook})
\[
 \rho = \Omega^2\,\hat{\rho} - \Omega = - \Omega + \ord{\Omega^3}.
\]

Let us now turn to the asymptotic behaviour of the matter fields. Appropriate conformal transformation of the four-potential $A_a$ is $\hat{A}_a = A_a$, so that the unphysical electromagnetic spinor $\hat{\phi}_{AB}$ is
\[
 \hat{\phi}_{AB} = \hat{\nabla}_{X'(A}\hat{A}{}^{X'}_{B)} = \Omega\,\phi_{AB}.
\]
Assuming that the unphysical quantities are of the order $\ord{1}$ near $\scri^+$, for the Newman-Penrose components of the potential we obtain
\begin{align}
 A_0 = A_al^a = \ord{\Omega^2}, \quad
 A_1= A_am^a = \ord{\Omega}, \quad
 A_{\bar{1}} = A_a\bar{m}^a = \ord{\Omega}, \quad
 A_2 = A_an^a = \ord{1}.
\end{align}
\noindent Similarly, for the electromagnetic spinor we find standard asymptotic behaviour in the form 
\begin{align}
\phi_0 = \phi_{AB}o^Ao^B =  \ord{\Omega^3}, \quad
\phi_1= \phi_{AB}o^A\iota^B =  \ord{\Omega^2}, \quad
\phi_2= \phi_{AB}\iota^A\iota^B  = \ord{\Omega}.
\end{align}

The spinor form of the Klein-Gordon equation (\ref{eq:nabla fii}) is genuinely not conformally-invariant and so we have to prescribe the conformal behaviour of the scalar field on the physical grounds. Natural requirement \cite{BST2} is that the physical scalar field vanishes at infinity, so we postulate
\begin{align}
 \phi = \Omega\,\hat{\phi} = \ord{\Omega},
\end{align}
assuming that $\hat{\phi}$ is regular on $\scri^+$. Components of the gradient $\fii_{a}=\nabla_a \phi$ then behave according to the formulae (recall (\ref{eq:NP operators asymp}) and (\ref{eq:CI PI asymp}))
\begin{align}
\begin{split} 
\fii_0 = \ord{\Omega^2}, \quad     
\fii_1 = \ord{\Omega^2}, \quad   
\fii_{\bar{1}} = \ord{\Omega^2}, \quad
\fii_2 = \ord{\Omega},
\end{split}
\end{align}
where the Newman-Penrose components of the field $\fii_a$ are defined by \eqref{eq:grad of phi}.

The Weyl spinor is conformally invariant with zero weight\footnote{This depends on the conventions used. What is convention-independent is the behaviour of the Weyl tensor $C_{abcd}=\Psi_{ABCD}\epsilon_{A'B'}\epsilon_{C'D'}$. In the non-abstract index formalism, components of tensors are related to components of spinors via van der Waerden symbols $\sigma_a^{AA'}$ which can have a conformal weight and thus they affect the conformal weight of $\Psi_{ABCD}$, as in, e.g. \cite{PenroseZRM}.}:
\[
 \Psi_{ABCD} = \hat{\Psi}_{ABCD}.
\]
Under certain weak assumptions it is possible to show \cite{StewartBook} that $\hat{\Psi}_{ABCD}$ vanishes on $\scri^+$ so that smoothness shows it is of order $\ord{\Omega}$. Hence, for the Weyl tensor we obtain usual asymptotic behaviour
\begin{align}
\Psi_0 = \ord{\Omega^5}, \quad
\Psi_1 = \ord{\Omega^4}, \quad
\Psi_2 = \ord{\Omega^3}, \quad
\Psi_3 = \ord{\Omega^2}, \quad
\Psi_4 = \ord{\Omega}.
\end{align}
Asymptotic behaviour of the components of the Ricci spinor can be found from Einstein's equations (\ref{eq:Ricci comps}):
\begin{align*}
\Phi_{00} &=  \ord{\Omega^4}, & \Phi_{01} &= \ord{\Omega^4}, & \Phi_{11} &= \ord{\Omega^2}, \\
\Phi_{02} &= \ord{\Omega^4},  & \Phi_{12} &= \ord{\Omega^3}, & \Phi_{22} &= \ord{\Omega^2}.
\end{align*}
Behaviour of the scalar curvature $\Lambda$ is found from \eqref{eq:Phi Lambda} to be
\begin{align}
 \Lambda &= \ord{\Omega^2}.
\end{align}
This completes the discussion of the conformal behaviour of physical and geometrical quantities used in the calculation.


\section{Asymptotic solution}
\label{sec:solution}

In this section we present the asymptotic solution of Einstein-Maxwell-Klein-Gordon equations introduced in the section \ref{sec:field eqs}. Let $X$ be any Newman-Penrose scalar quantity which is of the order $\ord{\Omega^n}$. Then, assuming analyticity of the solution, we expand this quantity into the series in coordinate $\Omega$ in the neighbourhood of $\scri$:
\begin{align}
 X = \sum_{k=0}^\infty X^{(k)}\,\Omega^{n+k}.
 \label{eq:X expansion}
\end{align}
Expanding all Newman-Penrose quantities\footnote{By the Newman-Penrose quantities we mean five components $\Psi_m$, $m=0,\dots 4$, six independent components $\Phi_{mn}$, $m,n=0,1,2$, twelve spin coefficients, three electromagnetic components $\phi_m$, $m=0,1,2$, four components of the potential $A_m$, $m=0,1,\bar{1},2$, and the scalar field $\phi$. } in this way and using the field equations we find the coefficients $X^{(0)}, X^{(1)},\dots$ in the leading terms of expansions \eqref{eq:X expansion}.

At the first stage we employ the Ricci identities \eqref{eq:Drho}, \eqref{eq:Dsigma}, \eqref{eq:Dalpha}, \eqref{eq:Dbeta} and \eqref{eq:deltarho-deltabarsigma} and the frame equation \eqref{eq:bardeltaPI-deltabarPI} which yield the following expansions of the spin coefficients $\rho, \sigma, \alpha$ and $\beta$:
\begin{subequations}
\label{eq:expansions}
\begin{align}
 \rho &= -\Omega - \left( \shear\shearcc+\phizero\phizerocc\right)\Omega^3 - \left(\phizero\,\bar{\phi}{}^{(1)} + \phi^{(1)}\,\phizerocc\right) \Omega^4 +  \ord{\Omega^5}, \\
 \sigma &= \phantom{-}\shear\,\Omega^2 + 
 \left( \shear{}^2\,\shearcc - \frac{1}{2}\,\Psi_0^{(0)} + \shear\,\phizero\,\phizerocc\right)\Omega^4+
 \ord{\Omega^5}, \\
 \alpha &= \phantom{-}a\,\Omega + \left(\eth\shearcc+a\shearcc\right)\Omega^2 + \ord{\Omega^3}, \\
 \beta &= -\,a\,\Omega - a\,\shear\,\Omega^2 + \ord{\Omega^3}, \\
 \pi =\bar{\tau} &= \phantom{-}(\eth\shearcc)\,\Omega^2 + \ord{\Omega^3},
\end{align}
\end{subequations}
where $\shear$ is the asymptotic shear of Newman and Penrose \cite{TimothyNewmanKozameh,NewmanPenrose} and 
\[
 a = - \frac{\cot\theta}{2\sqrt{2}}.
\]
Operators $\eth$ and $\bar{\eth}$ are defined by relations \cite{StewartBook}
\begin{align}
 \eth \eta = \hat{\delta}\eta + 2\,w\,a\,\eta, \qquad 
\bar{\eth}\eta = \hat{\bar{\delta}}\eta - 2\,w\,a\,\eta,
\end{align}
when acting on the scalar $\eta$ of the spin weight $w$.

Now, the $\ord{\Omega^2}$ terms in the Ricci identity \eqref{eq:Dmu-deltapi} give
\[
 m^2\,\phizero\,\phizerocc = 0,
\]
where $m$ is the mass of the scalar field. The coefficient $\phizero$ is the leading term in the asymptotic expansion of the scalar field and in fact represents the radiative component of the field. If we do not want to exclude the presence of the scalar radiation which is expected to contribute to the Bondi mass-loss formula, we are forced to set $m=0$. This is in agreement with the fact that massive fields do not extend to $\scri^+$, see \cite{Winicour,HelferMassiveFields,BST2}. Hence, in what follows we will consider only the massless scalar field.

Assuming now $m=0$ and $\phizero\neq 0$ and using all Ricci identities \eqref{eq:Drho}--\eqref{eq:deltarho-deltabarsigma} and the frame equations \eqref{eq:DH}--\eqref{eq:bardeltaPI-deltabarPI} we find the asymptotic expansion of remaining spin coefficients:

\begin{subequations}
\begin{align}
\lambda &= \phantom{-}\shearccdot\,\Omega + \left(\frac{1}{2}\shearcc-\bar{\eth}\eth\shearcc\right)\,\Omega^2 + \ord{\Omega^3}, \\
\mu &= -\frac{1}{2}\,\Omega - \left(
 \eth^2\shearcc +\shear\shearccdot + \Psi_2^{(0)} + \frac{1}{6}\pd_u(\phizero\phizerocc)\right)\Omega^2 + \ord{\Omega^3}, \label{eq:mu exp} \\
\gamma &= \phantom{-}\left( a\eth \shearcc - a\bar{\eth}\shear-\frac{1}{2}\Psi_2^{(0)}+\frac{1}{6}\pd_u(\phizero\bar{\phi}{}^0)\right)\Omega^2 + \ord{\Omega^3}, \\
\nu &= \ord{\Omega^2}.
\end{align}
\end{subequations}

Components of the metric tensor with respect to the coordinates $(u,r,\theta,\phi)$ are given in terms of the metric functions $H, C^I$ and $P^I$ satisfying the frame equations \eqref{eq:frame equations}. Their asymptotic expansions read
\begin{subequations}
\begin{align}
H &= -\frac{1}{2} + \left(  \frac{1}{3}\,\pd_u(\phizero\phizerocc) - \frac{1}{2}\Psi_2^{(0)}-\frac{1}{2}\bar{\Psi}_2^{(0)}\right)\Omega + \ord{\Omega^2},\\
C^2 &= - \frac{1}{\sqrt{2}}\left(\eth \shearcc + \bar{\eth}\shear\right)\,\Omega^2 + \ord{\Omega^3},\\
C^3 &= \phantom{-} \frac{i}{\sqrt{2}\sin\theta}\left(\eth \shearcc - \bar{\eth}\shear\right)\,\Omega^2 + \ord{\Omega^3}.
\end{align}
\end{subequations}

Similar expansions can be obtained for the components of the Ricci tensor and the Ricci scalar,
\begin{subequations}
\begin{align}
\Phi_{00} &= \phizero\,\phizerocc\,\Omega^4 + 2\,\left(\phi^{(1)}\,\phizerocc + \phizero\,\bar{\phi}^{(1)}\right) \,\Omega^5 + \ord{\Omega^6},\\
\Phi_{01} &= - \frac{1}{2}\eth(\phizero\,\phizerocc) \Omega^4 + \ord{\Omega^5}, \\
\Phi_{02} &= \left( -\,\phi_0^{(0)}\,\dot{A}_1^{(0)}+ (\eth\phizero+ieA_1^{(0)}\phizero)(\eth\phizerocc-ieA_1^{(0)}\phizerocc)\right)\Omega^4 + \ord{\Omega^5},\\
\Phi_{11} &= -\,\frac{1}{4}\pd_u\left(\phi^0\bar{\phi}{}^0\right)\Omega^3 + \ord{\Omega^4}, \\
\Phi_{12} &= \left( - \phi_1^{(0)}\,\dot{A}_1^{(0)} + \frac{1}{2}\,\dot{\bar{\phi}}{}^{(0)}(\eth\phizero+ieA_1^{(0)}\phizero)
+\frac{1}{2}\dot{\phi}{}^{(0)}(\eth\phizerocc - ie A_1^{(0)}\phizerocc)\right)\Omega^3 + \ord{\Omega^4},\\
\Phi_{22} &= \left( \dot{A}_1^{(0)}\,\dot{A}_{\bar{1}}^{(0)} + \dot{\phi}{}^{(0)}\,\dot{\bar{\phi}}{}^{(0)}\right) \Omega^2 + \ord{\Omega^3}, \\
\Lambda &= \phantom{-}\frac{1}{12}\pd_u\left(\phi^0\bar{\phi}{}^0\right)\Omega^3 + \ord{\Omega^4},
\end{align}
\end{subequations}
\noindent and for the components of the Weyl spinor,
\begin{subequations}
\begin{align}
\Psi_0 &= \Psi_0^{(0)}\,\Omega^5 + \Psi_0^{(1)}\,\Omega^6 + \ord{\Omega^7},\\
\Psi_1 &= \Psi_1^{(0)}\,\Omega^4 + \Psi_1^{(1)}\,\Omega^5 + \ord{\Omega^6},\\
\Psi_2 &= \Psi_2^{(0)}\,\Omega^3 + \ord{\Omega^4},\\
 \Psi_3 &=\Psi_3^{(0)}\,\Omega^2 + \ord{\Omega^3},\\
 \Psi_4 &= \Psi_4^{(0)}\,\Omega +\Psi_4^{(1)}\,\Omega^2 + \ord{\Omega^3},
\end{align}
\end{subequations}
where
\begin{subequations}
\begin{align}
\Psi_3^{(0)} &=  -\eth\shearccdot, \qquad 
\Psi_4^{(0)} = - \ddot{\bar{\sigma}}{}^0, \qquad 
 \Psi_4^{(1)} = \bar{\eth}{\eth}\dot{\bar{\sigma}}{}^0\label{eq:psi40}\\
\Psi_1^{(0)} &= -\,2\,\shear\,\eth\shearcc + 2\,a\,\shear\,\shearcc+(\eth + a)(\phizero\,\phizerocc), \\
\Psi_1^{(1)} &=  3\,\phi_0^{(0)}\,\bar{\phi}_1^{(0)} - \bar{\eth}\Psi_0^{(0)} - \shear \eth(\phi^{(0)}\,\bar{\phi}{}^{(0)}) +
\frac{1}{2}\,\left(\phi^{(0)}\,\eth\bar{\phi}{}^{(1)} + \bar{\phi}{}^{(0)}\,\eth\phi^{(1)}\right) + \frac{1}{2}\,\shear\bar{\eth}(\phi^{(0)}\,\bar{\phi}{}^{(0)})   \nonumber \\
& \phantom{=} - \left( \phi^{(1)}\,\eth\bar{\phi}{}^{(0)} + \bar{\phi}{}^{(1)}\,\eth\phi^{(0)}\right) 
+ \phi^{(0)}\,\bar{\phi}{}^{(0)} \left(
3\,e^2\,A_0^{(0)}\,A_1^{(0)} - \bar{\eth} \shear \right)  \nonumber\\
&\phantom{=} + \frac{3}{2}\,i\,e\left[
A_0^{(0)}\,\phi^{(0)} \left( \eth\bar{\phi}{}^{(0)}-\bar{\phi}{}^{(1)} \right)
-
A_1^{(0)} \,\bar{\phi}{}^{(0)} \left( \eth\phi^{(0)} - \phi^{(1)}\right)
\right], \\
\dot{\Psi}_2^{(0)} &= \frac{2}{3}\,\dot{\phi}{}^{(0)}\,\dot{\bar{\phi}}{}^{(0)} + \phi_2^{(0)}\,\bar{\phi}{}_2^{(0)} + \eth\Psi_3^{(0)}
-\frac{1}{6}\,\left(\ddot{\bar{\phi}}{}^{(0)}\,\phi^{(0)}+\ddot{\phi}{}^{(0)}\,\bar{\phi}{}^{(0)}\right) + \shear\,\Psi_4^{(0)}  + e^2\,(A_2^{(0)})^2\,\phi^{(0)}\,\bar{\phi}{}^{(0)} \nonumber \\
& \phantom{-} 
+ i\,e\,A_2{}^{(0)}\left(
\phi{}^{(0)}\,\dot{\bar{\phi}}{}^{(0)} - \dot{\phi}{}^{(0)}\,\bar{\phi}{}^{(0)}\right).
\label{eq:psi2dot}
\end{align}
\end{subequations}

For the components of electromagnetic spinor we find the following expansions:
\begin{subequations}
\begin{align}
\phi_0 &= \phi_0^{(0)}\,\Omega^3 + \phi_1^{(1)}\,\Omega^4 + \ord{\Omega^5}, \\
\phi_1 &= \phi_1^{(0)}\,\Omega^2 + \phi_1^{(1)}\,\Omega^3 + \ord{\Omega^4}, \\
\phi_2 &= \phi_2^{(0)}\,\Omega + \phi_2^{(1)}\,\Omega^2 + \ord{\Omega^3},
\end{align}
\end{subequations}
where
\begin{subequations}
\begin{align}
\phi_0^{(0)} &=  - \shear\,A_{\bar{1}}^{(0)} - \eth A_0^{(0)},\label{eq:phi00} \\
\phi_1^{(0)} &= - \eth A_{\bar{1}}^{(0)}, \\
\phi_2^{(0)} &= \eth A_2^{(0)} - \dot{A}_{\bar{1}}^{(0)}.\label{eq:phi20}
\end{align}
\end{subequations}

\section{Bondi mass}
\label{sec:bondi}

In this section we finally construct the expression for the Bondi mass. We adopt the approach based on the asymptotic twistor equation as described in \cite{StewartBook,TodHuggett}. The twistor equation reads
\begin{align}
 \nabla_{A'}{}^{(A} \omega^{B)} &= 0.
 \label{eq:twistor}
\end{align}
\noindent Spinor $\omega^A$ can be written as a linear combination of the basis spinors,
\[
 \omega^A = \omega^0\,o^A + \omega^1\,\iota^A.
\]
In the following we assume that the components
\[
 \omega^0 = - \iota_A\,\omega^A \quad \text{and} \quad \omega^1 = o_A \,\omega^A
\]
are regular on $\scri^+$. Null vector $m^a$ has the spin weight $1$ which, assuming $\epsilon_{AB}$ has the spin weight zero, implies that the spin weights of $o^A$ and $\iota^A$ are $1/2$ and $-1/2$, respectively. Consequently, the components $\omega^0$ and $\omega^1$ have spin weights $-1/2$ and $1/2$.

Twistor equation is conformally invariant if the spinor $\omega^A$ has conformal weight zero, i.e.
\[
 \omega^A = \hat{\omega}^A.
\]
In order to obtain explicit form of the twistor equation (\ref{eq:twistor}), we project it onto the spin basis and arrive at
\begin{subequations}
\begin{align}
D \omega^1 &= \phantom{-}\kappa\,\omega^0 + \eps \,\omega^1, & \Delta \omega^0 &= -\gamma\,\omega^0 - \nu\,\omega^1, \\
\bar{\delta}\omega^0 &= -\alpha\,\omega^0 - \lambda \,\omega^1, &\delta \omega^1 &=\phantom{-} \sigma\,\omega^0 + \beta\, \omega^1,  \\
D\omega^0 - \bar{\delta}\omega^1 &= -(\eps+\rho)\omega^0 - (\alpha+\pi)\omega^1, & \Delta\omega^1 - \delta \omega^0 &=\phantom{-} (\beta+\tau)\omega^0 + (\gamma+\mu)\omega^1.
\end{align}
\end{subequations}
\noindent In general spacetimes, these equations do not possess a non-trivial solution. Thus, since we are interested in the Bondi mass which is defined at null infinity, we restrict the twistor equation to $\scri$ in what follows.

Quantities $\omega^0$ and $\omega^1$ are regular by assumption and hence can be expanded in the neighbourhood of $\scri$ into the series of the form
\begin{align}
\omega^0 = \omega^0_0 + \omega^0_1\,\Omega + \ord{\Omega^2}, \qquad
\omega^1 = \omega^1_0 + \omega^1_1\,\Omega + \ord{\Omega^2}.
\label{eq:expansions omega}
\end{align}
\noindent Using expansions of the spin coefficients and the Newman-Penrose operators, we find that leading terms $\omega^0_0$ and $\omega^1_0$ satisfy relations
\begin{align}
 \eth \omega^1_0 = 0, \qquad \bar{\eth}\omega^1_0 = -\omega^0_0, \qquad \dot{\omega}{}^1_0 = 0,
 \qquad
 \omega^1_1 = 0,
 \label{eq:restricted twistor eq}
\end{align}
where the dot denotes differentiation with respect to the variable $u$. 

Next we define the symmetric spinor \cite{Szabados2dSenInGR,Szabados2dSenquasi}
\begin{align}
u_{AB} &= \frac{1}{2}\left( \omega_{(A}\nabla^{C'}_{B)}\bar{\omega}_{C'} - \bar{\omega}_{C'} \nabla^{C'}_{(A} \omega_{B)} \right),
\end{align}
and the associated two form
\begin{align}
\FF_{ab} &= u_{AB}\,\epsilon_{A'B'} + \bar{u}_{A'B'}\,\epsilon_{AB}.
\end{align}
Now, following \cite{StewartBook}, we choose a null hypersurface $\Sigma$ which extends to $\scri^+$ and define $S(\Omega)$ to be the two surface $\Omega=\text{constant}$ in $\Sigma$. Hence, the hypersurface $\Sigma$ intersects $\scri^+$ at the two sphere $S(0)$. In addition, we define
\[
 I(\Omega) = \oint_{S(\Omega)} \FF_{ab}\,l^a n^b\,\dd S
\]
and
\begin{align}
 I_0 &= \lim_{\Omega \to 0} I(\Omega)
\label{eq:I0 def}
 \end{align}
if the limit exists. The Bondi four-momentum $P^a$ is then defined by the equation
\[
 I_0 = P^a k_a,
\]
where $k^a = \omega^A \bar{\omega}{}^{A'}$.

The induced volume form on the two surface $S(\Omega)$ is 
\[
 {}^{(2)}\epsilon_{cd} = n^a l^b \epsilon_{abcd} = i\left( \epsilon_{C'D'} o_C \iota_D - \epsilon_{CD} \bar{o}_{C'}
\bar{\iota}_{D'}\right) = \ord{\Omega^{-2}}.
\]
Thus, in order to show that the limit $I_0$ exists we have to show that the integrand behaves as
\[
 \FF_{ab} l^a n^b = \ord{\Omega^2}.
\]
Direct calculation shows
\[
 \FF_{ab}l^a n^b = \rho\,\omega^0\,\bar{\omega}^0 + \mu\,\omega^1\,\bar{\omega}^1 + \Re\left(\pi\,\omega^1\,\bar{\omega}^0 + \bar{\omega}^1\,\delta\omega^0 - \bar{\omega}^0\bar{\delta}\omega^1\right).
\]
Using expansions \eqref{eq:expansions} and \eqref{eq:expansions omega} we find
\begin{align}
\FF_{ab}l^a n^b = \Omega \,\Re \left[ - \omega^0_0\,\bar{\omega}^0_0 -\frac{1}{2}\,\omega^1_0\,\bar{\omega}^1_0 + 
\bar{\omega}^1_0 \eth\omega^0_0 - \bar{\omega}^0_0 \bar{\eth}\omega^1_0
\right] + \ord{\Omega^2}.
\end{align}
By \eqref{eq:restricted twistor eq} we have
\[
- \bar{\omega}^0_0\,\bar{\eth}\omega^1_0 =  \omega^0_0\,\bar{\omega}^0_0
\]
and so
\begin{align}
\FF_{ab}l^a n^b = \Omega \,\Re \left[ -\frac{1}{2}\,\omega^1_0\,\bar{\omega}^1_0 + 
\bar{\omega}^1_0 \eth\omega^0_0 \right] + \ord{\Omega^2}.
\end{align}
Using the commutator
\[
[\eth,\bar{\eth}]\omega^1_0 = -\frac{1}{2}\,\omega^1_0
\]
and asymptotic twistor equation \eqref{eq:restricted twistor eq} we find
\[
\eth\omega^0_0 = - \eth\bar{\eth}\omega^1_0 = - \bar{\eth}\eth\omega^1_0 + \frac{1}{2} \omega^1_0
\]
which implies
\[
 \FF_{ab}l^a n^b = \ord{\Omega^2}
\]
and hence the limit $I_0$ in \eqref{eq:I0 def} exists.

Expanding the quantity $\FF_{ab}l^an^b$ further we arrive at
\begin{multline}
\FF_{ab}l^a n^b = \Omega^2\,\Re \left[
- 2\,\omega^0_0\,\bar{\omega}^0_1 - \omega^1_0\,\bar{\omega}^1_1 + \mu^1\,\omega^1_0\,\bar{\omega}^1_0 +
(\eth\shearcc)\,\omega^1_0\,\bar{\omega}^0_0  + \shearcc\,\bar{\omega}^0_0\,\eth\omega^1_0
\right. \\
\left.
- \shear\,\bar{\omega}^1_0\,\bar{\eth}\omega^0_0
+\bar{\omega}^1_1\,\eth\omega^0_0 + \bar{\omega}^1_0\,\eth\omega^0_1 - \bar{\omega}^0_1 \,\bar{\eth}\omega^1_0
- \bar{\omega}^0_0 \,\bar{\eth}\omega^1_1
\right] + \ord{\Omega^3}.
\end{multline}
Imposing \eqref{eq:restricted twistor eq} this simplifies to
\begin{align}
\begin{split}
\FF_{ab}l^a n^b {}={}& \Omega^2\,\Re \left[
- \omega^0_0\,\bar{\omega}^0_1  + \mu^1\,\omega^1_0\,\bar{\omega}^1_0 +
(\eth\shearcc)\,\omega^1_0\,\bar{\omega}^0_0  
- \shear\,\bar{\omega}^1_0\,\bar{\eth}\omega^0_0
 + \bar{\omega}^1_0\,\eth\omega^0_1 
\right]  + \ord{\Omega^3}.
\end{split}
\end{align}
Next we have
\[
 \Re\left[ 
 -\omega^0_0\,\bar{\omega}^0_1 + \bar{\omega}^1_0\eth\omega^0_1
 \right] =
 \Re\left[
 (\bar{\eth}\omega^1_0)\bar{\omega}^0_1 + \bar{\omega}^1_0\eth\omega^0_1
 \right] = 
 \Re\left[
 \omega^1_0\eth\bar{\omega}^1_0 + \bar{\omega}^1_0\eth\omega^0_1
 \right] = 
 \Re \left[ \eth (\omega^0_1 \bar{\omega}^1_0)\right]
\]
which vanishes on integration,
\[
 \oint \eth (\omega^0_1 \bar{\omega}^1_0) \,\dd\hat{S} = 0,
\]
because quantity $\omega^0_1 \bar{\omega}^1_0$ has the spin weight $-1$. Thus,
\begin{align}
 I_0 &= \oint \Re\left[ \mu^1 \omega^1_0 \bar{\omega}^1_0 - \shear \,\bar{\omega}^1_0 \bar{\eth}\omega^0_0 
 +
 \omega^1_0 \bar{\omega}^0_0 \eth\shearcc
 \right]\,\dd\hat{S}.
 \label{eq:I0 temp}
\end{align}

Let us use equation \eqref{eq:restricted twistor eq} again to rearrange the third term of the integrand \eqref{eq:I0 temp},
\[
 \oint \omega^1\,\bar{\omega}^0_0\,\eth\shearcc \,\dd\hat{S} = 
 - \oint \eth\left(\omega^1_0\,\shearcc\,\eth\bar{\omega}^1_0 \right)\dd\hat{S} +
 \oint \shearcc\, \omega^1_0\,\eth \left(\eth\bar{\omega}^1_0\right)\dd\hat{S},
\]
where the first integral on the right hand side vanishes because of the spin weight of the argument of the $\eth$ operator. Next we expand quantity $\omega^1_0$ of the spin weight $1/2$ into the series in spin-weighted spherical harmonics,
\[
 \omega^1_0 = \sum_{l=0}^\infty\sum_{m=-l-1}^l a_{lm}\,\Yslm{\frac{1}{2}}{l+\frac{1}{2},~}{m+\frac{1}{2}},
\]
where the coefficients $a_{lm}$are time-independent by \eqref{eq:restricted twistor eq}. Since the operator $\eth$ ($\bar{\eth}$) acts as the spin raising (lowering) operator, we can write
\[
 \eth \Yslm{s}{l}{m} = c_{slm}\,\Yslm{s+1}{l}{m},\qquad
 \bar{\eth} \Yslm{s}{l}{m} = d_{slm}\,\Yslm{s-1}{l}{m},
\]
where particular form of coefficients $c_{slm}$ and $d_{slm}$ is not important. Applying $\eth$ on $\omega^1_0$ and imposing \eqref{eq:restricted twistor eq} yields
\[
 \eth \omega^1_0 = \sum_{l=0}^\infty\;\;\sum_{m=-l-1}^l a_{lm}\,c_{\frac{1}{2}\,lm}\,\Yslm{\frac{3}{2}}{l+\frac{1}{2},~}{m+\frac{1}{2}} = 0.
\]
Functions $\Yslm{\frac{3}{2}}{\frac{1}{2}}{m}$ vanish by definition while  the orthogonality of spin-weighted spherical harmonics implies
\[
 a_{lm} = 0 \quad \text{for} \quad l > 0.
\]
The quantity $\omega^1_0$ then acquires the form
\begin{align}
 \omega^1_0 = a\,\Yslm{\frac{1}{2}}{\frac{1}{2},~}{-\frac{1}{2}} + b\,\Yslm{\frac{1}{2}}{\frac{1}{2},~}{\frac{1}{2}}.
\label{eq:omega10 solution}
\end{align}
Application of $\bar{\eth}{}^2$ to this expansion immediately yields
\begin{align}
 \bar{\eth}^2 \omega^1_0 = \tilde{a}\,\Yslm{-\frac{3}{2}}{\frac{1}{2},~}{-\frac{1}{2}}
 +\tilde{b}\,\Yslm{-\frac{3}{2}}{\frac{1}{2},~}{\frac{1}{2}} = 0.
 \label{eq:ethbar^2 omega^1_0}
\end{align}
Hence, 
\[
  \oint \omega^1_0\,\bar{\omega}^0_0\,\eth\bar{s} \,\dd\hat{S} = 0.
\]
Finally, the last vanishing term in \eqref{eq:I0 temp} is
\[
 \oint \shear\,\bar{\omega}^1_0\,\bar{\eth}\omega^0_0\dd\hat{S} =
 -\oint \shear\,\bar{\omega}^1_0\,\bar{\eth}^2\omega^1_0\dd\hat{S} = 0
\]
by \eqref{eq:restricted twistor eq} and \eqref{eq:ethbar^2 omega^1_0}.

Thus, we have found that the integral $I_0$ exists and reduces to
\begin{align}
I_0 &= \Re\oint \mu^1\,\omega^1_0\,\bar{\omega}{}^1_0\dd\hat{S},
 \label{eq:I0 solution}
\end{align}
\noindent where $\mu^1$ is $\ord{\Omega^2}$ term in \eqref{eq:mu exp} so that the integral reads
\begin{align}
I_0 &= \oint \left( \Psi_2^0 + \shear\,\shearccdot + \frac{1}{6}\,\frac{\pd}{\pd u} (\phi^0\,\bar{\phi}{}^0)\right)
\omega^1_0\,\bar{\omega}{}^1_0\,
\dd\hat{S},
\end{align}
where we have used
\[
 \oint \eth^2\bar{s} \,\dd\hat{S}  = 0.
\]

Now, \eqref{eq:omega10 solution} implies that the spin weight zero quantity $\omega^1_0\bar{\omega}^1_0$ can be expanded as
\begin{align}
 \omega^1_0\,\bar{\omega}^1_0 &= \alpha\,\Ylm{0}{0} + \sum_{m=1}^1 \beta_m\,\Ylm{1}{m}
\end{align}
for some coefficients $\alpha$ and $\beta_m$. Since the Bondi mass is a zeroth component of the four-momentum, we set $\beta_m = 0$ and $\alpha=1$ which corresponds to the timelike direction. With this choice, we arrive at the expression for the Bondi mass of electro-scalar spacetimes in the form
\begin{align}
M_B &= \frac{1}{2\sqrt{\pi}} \oint \left( \Psi_2^0 + \shear\,\shearccdot + \frac{1}{6}\,\frac{\pd}{\pd u} (\phi^0\,\bar{\phi}{}^0)\right)\dd\hat{S}.
\label{eq:Bondi mass}
\end{align}
Corresponding mass-loss formula is found by taking the derivative of \eqref{eq:Bondi mass} with respect to variable $u$ and using equations \eqref{eq:psi2dot} and \eqref{eq:psi40}:
\begin{align}
 \dot{M}_B = -\frac{1}{2\sqrt{\pi}}\oint \left[
\sheardot\,\shearccdot+ \phi_2^0\,\bar{\phi}{}_2^0 + \dot{\phi}{}^0\,\dot{\bar{\phi}}{}^0
+ i\,e\,A_2^0\left(\phi^0\,\dot{\bar{\phi}}{}^0
-\dot{\phi}{}^0\,\bar{\phi}{}^0\right)
+ e^2\,A_2^0\,A_2^0\,\phi^0\,\bar{\phi}{}^0
 \right]\dd \hat{S}.
 \label{eq:Bondi massloss}
\end{align}
Clearly, the first three terms represent the mass-loss by gravitational, electromagnetic and scalar radiation, while remaining terms represent the mass-loss by interactions between electromagnetic and scalar fields.

The Bondi mass-loss formula can be brought into simpler form when we define
\[
 \DD_u \phizero= \pd_u\phizero + i\,e\,A_2^0\,\phizero, \qquad
\DD_u \phizerocc = \pd_u \phizerocc - i\,e\,A_2^0\,\phizerocc,
\]
so that $\DD_u$ is the projection of the gauge covariant derivative $n^a\DD_a$ restricted to $\scri$. In terms of the operator $\DD_u$, the Bondi mass-loss formula reads
\begin{align}
 \dot{M}_B = -\frac{1}{2\sqrt{\pi}}\oint \left[
\sheardot\,\shearccdot+ \phi_2^0\,\bar{\phi}{}_2^0 + 
\left(\DD_u \phizero\right)\left(\DD_u \phizerocc\right)
 \right]\dd \hat{S}.
 \label{eq:Bondi massloss covariant}
\end{align}
This expression is manifestly {\itshape gauge invariant} and {\itshape negative semi-definite}. Hence, unlike the conformal scalar field with indefinite ``mass-loss'' formula \eqref{eq:Bondi massloss indefinite}, in the case of interacting electromagnetic and massless Klein-Gordon fields, the Bondi mass is either constant or decreasing function of time. Alternatively, expression \eqref{eq:Bondi massloss covariant} can be rewritten in terms of the four-potential $A_a$ using the relation \eqref{eq:phi20}:
\begin{align}
 \dot{M}_B = -\frac{1}{2\sqrt{\pi}}\oint \left[
\sheardot\,\shearccdot+\dot{A}_1^{(0)}\,\dot{A}_{\bar{1}}^{(0)} + \dot{\phi}{}^{(0)}\,\dot{\bar{\phi}}{}^{(0)}
 \right]\dd \hat{S}.
 \label{eq:Bondi massloss not-covariant}
\end{align}

In the absence of electromagnetic field, formulae \eqref{eq:Bondi mass} and \eqref{eq:Bondi massloss covariant} reduce to expressions \eqref{eq:Bondi mass BST2} and \eqref{eq:Bondi massloss BST2} found in \cite{BST2} for the massless Klein-Gordon field. 

\section{Conclusion}

In this paper we have derived the spinor equations for the system of coupled gravitational, electromagnetic and scalar fields and found the asymptotic solution of this system in the neighbourhood of the future null infinity. The asymptotic solution reduces to the well-known expansions for electrovacuum spacetimes \cite{PenroseRindlerII,NewmanPenrose} and our previous results on spacetimes with the scalar field sources \cite{BST2}. Using this solution and the solution of asymptotic twistor equation, we have arrived at the expression for the Bondi mass of resulting electro-scalar spacetime, equation \eqref{eq:Bondi mass}. This expression coincides with \eqref{eq:Bondi mass BST2}. 

The Bondi mass-loss formula has been derived and expressed in terms of the four-potential \eqref{eq:Bondi massloss not-covariant} and in the gauge invariant form \eqref{eq:Bondi massloss covariant} which is manifestly negative semi-definite. This last result shows that in the case of electro-scalar spacetimes, the Bondi mass is a non-increasing function of time.


\begin{acknowledgements}
This work was supported by the grant GAUK 606412 of the Charles University in Prague, Czech Republic. The authors acknowledge useful discussions with Paul Tod and Szabados L\'aszl\'o. In particular, we are grateful to Paul Tod for his comments on the original manuscript. 
\end{acknowledgements}


\appendix


\section{Field equations in the Newman-Penrose formalism}

\label{app:projections}

Four-potential $A_a = A_{AA'}$ is a real vector field and its components with respect to the spin basis will be denoted by
\begin{subequations}
\begin{align}
A_0 &= A_{XX'}o^X\bar{o}^{X'}, &
A_1 &= A_{XX'}o^X\bar{\iota}^{X'}, \\
A_{\bar{1}} &= A_{XX'} \iota^X \bar{o}^{X'}, &
A_2 &= A_{XX'}\iota^X\bar{\iota}^{X'}.
\label{eq:A_a components}
\end{align}
\end{subequations}
Similarly we introduce the Newman-Penrose components of electromagnetic spinor $\phi_{AB}$ by
\begin{align}
 \phi_0 = \phi_{AB}\,o^A\,o^B, \qquad 
 \phi_1 = \phi_{AB}\,o^A\,\iota^B, \qquad
 \phi_2 = \phi_{AB}\,\iota^A\,\iota^B.
\end{align}
Potential $A_a$ is governed by equation (\ref{eq:nabla A}),
\[
\nabla^{A'}_A A_{BA'}  =  -\,\phi_{AB}.
\]
Projections of this equation onto the spin basis are
\begin{subequations}
\begin{align}
DA_1-\delta A_0 &= (\bar{\pi}-\bar{\alpha}-\beta)A_0 + (\eps-\bar{\eps}+\bar{\rho})A_1+\sigma A_{\bar{1}} - \kappa A_2+\phi_0, \label{eq:DA1-deltaA0} \\
DA_2-\delta A_{\bar{1}} &= -\mu A_0 +\pi A_1 +(\bar{\pi}-\bar{\alpha}+\beta)A_{\bar{1}}+(\bar{\rho}-\eps-\bar{\eps})A_2+\phi_1,
\label{eq:DA2-deltaA1bar} \\
\Delta A_0-\bar{\delta}A_1 &= (\gamma+\bar{\gamma}-\bar{\mu})A_0 + (\bar{\beta}-\alpha-\bar{\tau})A_1 - \tau A_{\bar{1}} + \rho A_2 - \phi_1,
\label{eq:DeltaA0-bardeltaA1}\\
\Delta A_{\bar{1}}-\bar{\delta}A_2 &= \nu A_0 - \lambda A_1 + (\bar{\gamma}-\gamma-\bar{\mu})A_{\bar{1}} + (\alpha+\bar{\beta}-\bar{\tau})A_2-\phi_2.
\label{eq:DeltaA1bar-bardeltaA2}
\end{align}
\label{eq:Nabla A NP}
\end{subequations}
The Lorenz condition $\nabla^a A_a = 0$ in the Newman-Penrose formalism acquires the form
\begin{align}
DA_2 - \Delta A_0 - \delta A_{\bar{1}} - \bar{\delta}A_1 &= (\gamma+\bar{\gamma}-\mu-\bar{\mu})A_0 +
(\pi-\alpha+\bar{\beta}-\bar{\tau})A_1 + (\bar{\pi}-\bar{\alpha}+\beta-\tau)A_{\bar{1}} + (\rho+\bar{\rho}-\eps-\bar{\eps})A_2 = 0.
\end{align}

Projections of the gradient $\fii_{AA'}=\nabla_{AA'}\phi$ will be denoted by
\begin{align}
\begin{split}
\fii_0 = D\phi, \qquad
\fii_2 = \Delta\phi, \qquad
\fii_1 = \delta\phi, \qquad
\fii_{\bar{1}} = \bar{\delta}\phi, \\
\bar{\fii}_0 = D\bar{\phi}, \qquad
\bar{\fii}_2 = \Delta\bar{\phi}, \qquad
\bar{\fii}_1 = \delta\bar{\phi},\qquad
\bar{\fii}_{\bar{1}} = \bar{\delta}\,\bar{\phi}
\end{split}
\label{eq:grad of phi}
\end{align}

Now we can complete equations for electromagnetic field. Equation (\ref{eq:nabla phiAB}),
\begin{align*}
\nabla^A_{B'}\phi_{AB} &= \frac{ie}{2}\left(
\bar{\phi}\,\fii_b - \phi\,\bar{\fii}_b
\right) - e^2\,\phi\,\bar{\phi}\,A_b,
\end{align*}
\noindent is the spinor version of Maxwell's equations with four-current $j^a$ on the right hand side. Projections of this equation onto the spin basis follow:

\begin{subequations}
\begin{align}
D\phi_1-\bar{\delta}\phi_0 &= 
(\pi-2\alpha)\phi_0+2\rho\phi_1-\kappa\phi_2+\frac{ie}{2} \left(\phi\bar{\fii}_0-\bar{\phi}\fii_0\right)+e^2\phi\bar{\phi}A_0,
\label{eq:Dphi1-bardeltaphi0} \\
D\phi_2-\bar{\delta}\phi_1 &= 
-\lambda\phi_0+2\pi\phi_1+(\rho-2\eps)\phi_2+\frac{ie}{2}\left(\phi\bar{\fii}_{\bar{1}}-\bar{\phi}\fii_{\bar{1}}\right)+e^2\phi\bar{\phi}A_{\bar{1}},
\label{eq:Dphi2-bardeltaphi2} \\
\Delta\phi_0-\delta\phi_1 &= (2\gamma-\mu)\phi_0 - 2\tau\phi_1+\sigma\phi_2 +\frac{ie}{2}\left( \bar{\phi}\fii_1-\phi\bar{\fii}_1\right)-e^2\phi\bar{\phi}A_1,
\label{eq:Deltaphi0-deltaphi1}\\
\Delta\phi_1-\delta\phi_2 &= \nu\phi_0 - 2\mu\phi_1 +(2\beta-\tau)\phi_2+\frac{ie}{2}\left(\bar{\phi}\fii_2-\phi\bar{\fii}_2\right)-e^2\phi\bar{\phi}A_2.
\label{eq:Deltaphi1-deltaphi2}
\end{align}
\label{eq:Maxwell eqs}
\end{subequations}

\noindent Dynamical equation for the gradient $\fii_{AA'}$ is provided by equation (\ref{eq:nabla fii})
\begin{align}
\nabla^A_{A'}\fii_{AB'} &= \,i\,e\,A^c\fii_c\,\epsilon_{A'B'} + \frac{1}{2}\left(m^2 - e^2\,A^c A_c\right)\phi\,\epsilon_{A'B'}.
\end{align}
\noindent Projected on the spin basis, this equation is equivalent to any of the following four scalar equations:
\begin{subequations}
\begin{align}
D\fii_{\bar{1}}-\bar{\delta}\fii_0 &= (\pi-\alpha-\bar{\beta})\fii_0+\bar{\sigma}\fii_1+(\rho+\bar{\eps}-\eps)\fii_{\bar{1}}-\bar{\kappa}\fii_2,
\label{eq:Dfiibar1-bardeltafii0}\\
D\fii_2-\bar{\delta}\fii_1 &= -\bar{\mu}\fii_0+(\pi-\alpha+\bar{\beta})\fii_1+\bar{\pi}\fii_{\bar{1}}+(\rho-\eps-\bar{\eps})\fii_2 -\phi \,m^2/2 +e^2\phi\left(A_0 A_2-A_1 A_{\bar{1}}\right) +
ie\left(
A_1\fii_{\bar{1}}+A_{\bar{1}}\fii_1-A_0\fii_2-A_2\fii_0
\right),
\label{eq:Dfii2-bardeltafii1} \\
\Delta\fii_0-\delta\fii_{\bar{1}} &= (\gamma+\bar{\gamma}-\mu)\fii_0 -\bar{\tau}\fii_1+(\beta-\bar{\alpha}-\tau)\fii_{\bar{1}}+\bar{\rho}\fii_2 -\phi\, m^2/2+e^2\phi\left(A_0 A_2-A_1 A_{\bar{1}}\right) +
ie\left(
A_1\fii_{\bar{1}}+A_{\bar{1}}\fii_1-A_0\fii_2-A_2\fii_0
\right),
\label{eq:Deltafii0-deltafiibar1} \\
\Delta\fii_1-\delta\fii_2 &= \bar{\nu}\fii_0+(\gamma-\bar{\gamma}-\mu)\fii_1-\bar{\lambda}\fii_{\bar{1}}+(\bar{\alpha}+\beta-\tau)\fii_2.
\label{eq:Deltafii1-deltafii2}
\end{align}
\label{eq:Nabla fii}
\end{subequations}

The Ricci spinor is related to the electro-scalar fields by Einstein's equations \eqref{eq:Einsteins eqs} and is given by formula \eqref{eq:Phi Lambda}. The Newman-Penrose components of the Ricci spinor read:

\begin{subequations}
\begin{align}
\Phi_{00} &= \phi_0\,\bar{\phi}_0 + \left(\DD_0 \phi\right)\left(\DD_0\bar{\phi}\right)  =  \phi_0\bar{\phi}_0 + \fii_0\bar{\fii}_0 + e^2 A_0^2 \phi\bar{\phi}+ieA_0\left( \phi\bar{\fii}_0-\bar{\phi}\fii_0\right),  \\
\Phi_{01} &=  \phi_0\,\bar{\phi}_1 + \left(\DD_{(0} \phi\right)\left(\DD_{1)}\bar{\phi}\right)=\phi_0\bar{\phi}_1 + \fii_{(0}\bar{\fii}_{1)} + e^2 \phi\bar{\phi} A_0 A_1 + ie\phi A_{(0}\bar{\fii}_{1)} - ie\bar{\phi} A_{(0}\fii_{1)},\\
\Phi_{11} &= \phi_1\,\bar{\phi}_1 + \frac{1}{2}\,\left[ 
\left(\DD_{(0}\phi\right)\left(\DD_{2)}\bar{\phi}\right) +
\left(\DD_{(1}\phi\right)\left(\DD_{\bar{1})}\bar{\phi}\right)
\right] \\
&= \phi_1\,\bar{\phi}_1 + \frac{1}{2}\left[
\fii_{(0}\bar{\fii}_{2)}  + \fii_{(1}\bar{\fii}_{\bar{1})} 
+ ie\phi \left(
A_{(0}\bar{\fii}_{2)} + A_{(1}\bar{\fii}_{\bar{1})}  
\right)
- ie\bar{\phi} \left(
A_{(0}{\fii}_{2)} + A_{(1}{\fii}_{\bar{1})}  
\right)
+ e^2 \phi \bar{\phi} \left(
A_0 A_2 - A_1 A_{\bar{1}}
\right)
\right],\\
\Phi_{02} &= \phi_0 \bar{\phi}_2 + \left(\DD_1\phi\right)\left( \DD_{1}\bar{\phi}\right) 
= 
\phi_0 \bar{\phi}_2 + \fii_{1}\bar{\fii}_{1} + e^2 \phi\bar{\phi}A_1^2 +ie\left(\phi A_1 \bar{\fii}_1 - \bar{\phi}A_1\fii_1\right), \\
\Phi_{12} &= \phi_1 \bar{\phi}_2 + \left(\DD_{(1}\phi\right)\left( \DD_{2)}\bar{\phi}\right) 
= \phi_1\bar{\phi}_2 + \fii_{(1}\bar{\fii}_{2)} + e^2 \phi\bar{\phi} A_1 A_2 
+ ie\left(\phi A_{(2}\bar{\fii}_{1)}-\bar{\phi}A_{(2}\fii_{1)}\right),\\
\Phi_{22} &= \phi_2\bar{\phi}_2 + \left(\DD_{2}\phi\right)\left( \DD_{2}\bar{\phi}\right) 
=\phi_2\bar{\phi}_2 + \fii_2\,\bar{\fii}_2 +  e^2\phi\bar{\phi} A_2^2 +ieA_2\left(\phi\bar{\fii}_2-\bar{\phi}\fii_2\right).
\end{align}
\label{eq:Ricci comps}
\end{subequations}

\begin{align}
6 \Lambda &=   \fii_{(1}\bar{\fii}_{\bar{1})} - \fii_{(0}\bar{\fii}_{2)}
 + i e \bar{\phi}\left(A_{(0}\fii_{2)} - A_{(1}\fii_{\bar{1})}\right) 
 + i e \phi\left( A_{(1}\bar{\fii}_{\bar{1})} - A_{(0}\bar{\fii}_{2)}\right)
+e^2\phi\bar{\phi}\left( A_{\bar{1}}A_1-A_0 A_2 \right) 
+ m^2\phi\bar{\phi}.
\label{eq:Lambda NP}
\end{align}

The Ricci identities in the tetrad introduced in section \ref{sec:asymptotics} simplify to the following set of equations.

\begin{subequations}
\begin{align}
D \rho &= \rho^2 + \sigma\,\bar{\sigma} + \Phi_{00}, \label{eq:Drho} \\
D\sigma &= 2\,\rho\,\sigma + \Psi_0, \label{eq:Dsigma} \\
D\alpha &= \rho\,\alpha + \beta\,\bar{\sigma} + \rho\,\pi + \Phi_{10}, \label{eq:Dalpha}, \\
D\beta &= (\alpha+\pi)\,\sigma + \rho\,\beta + \Psi_1, \label{eq:Dbeta}\\
D\gamma &= 2\,\bar{\pi}\,\alpha + 2\,\pi\,\beta + \pi\,\bar{\pi} + \Psi_2 - \Lambda + \Phi_{11}, \label{eq:Dgamma} \\
D\lambda - \bar{\delta}\pi &= \rho\,\lambda + \mu\,\bar{\sigma} + 2\,\alpha\,\pi + \Phi_{20}, \label{eq:Dlambda-bardeltapi}\\
D\mu - \delta\pi &= \rho\,\mu + \sigma\,\lambda + 2\,\beta\,\pi+ \Psi_2 + 2\,\Lambda, \label{eq:Dmu-deltapi} \\
D\nu - \Delta\pi &= 2\,\pi\,\mu + 2\,\bar{\pi}\,\lambda + (\gamma-\bar{\gamma})\pi + \Psi_3 + \Phi_{21}, \label{eq:Dnu-Deltapi}\\
D\tau &= 2\,\bar{\pi}\,\rho + 2\,\pi\,\sigma + \Psi_1 + \Phi_{01}, \label{eq:Dtau}\\
\Delta\rho - \bar{\delta}\tau &= (\gamma+\bar{\gamma}-\mu)\rho - \sigma\,\lambda - 2\,\alpha\,\tau - \Psi_2 - 2\,\Lambda, \label{eq:Deltarho} \\
\Delta\sigma - \delta\tau &= - (\mu-3\,\gamma+\bar{\gamma})\,\sigma - \bar{\lambda}\,\rho - 2\,\beta\,\tau - \Phi_{02}, \label{eq:Deltasigma} \\
\Delta\lambda - \bar{\delta}\nu &= -(2\,\mu + 3\,\gamma - \bar{\gamma})\,\lambda - (3\,\alpha+\beta)\,\nu, \label{eq:Deltalambda}\\
\Delta\alpha - \bar{\delta}\gamma &= \rho\,\nu - (\beta+\tau)\,\lambda + (\bar{\gamma}-\mu)\,\alpha + (\bar{\beta}-\bar{\tau})\,\gamma - \Psi_3, \label{eq:Deltaalpha} \\
\Delta\beta - \delta\gamma &= -\,\mu\,\tau + \sigma\,\nu + (\gamma-\bar{\gamma}-\mu)\,\beta - \alpha\,\bar{\lambda} - \Phi_{12}, \label{eq:Deltabeta} \\
\Delta\mu - \delta\nu &= - (\mu+\gamma+\bar{\gamma})\,\mu - \lambda\,\bar\lambda + \bar{\nu}\,\pi + 2\,\beta\,\nu - \Phi_{22}, \label{eq:Deltamu}\\
\delta\alpha - \bar{\delta}\beta &= \mu\,\rho- \lambda\,\sigma + \alpha\,\bar{\alpha} + \beta\,\bar{\beta} - 2\,\alpha\,\beta - \Psi_2 + \Lambda + \Phi_{11}, \label{eq:deltaalpha} \\
\delta\lambda - \bar{\delta}\mu &= \pi\,\mu + (\bar{\alpha}-3\,\beta)\,\lambda - \Psi_3 + \Phi_{21}, \label{eq:deltalambda},\\
\delta\rho-\bar{\delta}\sigma &= (\bar{\alpha} + \beta)\,\rho - (3\,\alpha-\bar{\beta})\,\sigma - \Psi_1 + \Phi_{01}. \label{eq:deltarho-deltabarsigma}
\end{align}
\end{subequations}

\bibliographystyle{spmpsci}      
\bibliography{bibliography}   

\end{document}